\documentclass[twocolumn,prc,aps,superscriptaddress,showpacs]{revtex4}

\usepackage{graphicx}
\usepackage{amssymb}
\usepackage[latin2]{inputenc}
\usepackage[T1]{fontenc}
\begin{document}

\title{Radiative electron capture in the first
       forbidden unique decay of $^{81}$Kr}

\date{\today}

%%%%%%%%%%%%%%%%%% introduction %%%%%%%%%%%%%%%%%%%%%%%%
\author{S.~Mianowski}
\author{E.~Werner-Malento}
\altaffiliation{Present Address: Institute of Physics, Polish Academy of Science,
Al. Lotnik\'ow 32/46, 02-668, Warsaw, Poland}
%
%\author{Z.~Janas}
%
\author{A.~Korgul}
\author{M. Pomorski}
\author{K. Pachucki}
\author{M.~Pf\"utzner}
\email{pfutzner@fuw.edu.pl}
\author{B. Szweryn}
\author{J. \.Zylicz}
\affiliation{Physics Department, University of Warsaw, Ho\.za 69, 00-681 Warsaw, Poland}
\author{P. Hornsh\o j}
\affiliation{Institute of Physics and Astronomy, University of Aarhus, DK-8000 Aarhus C, Denmark}
\author{T. Nilsson}
\affiliation{Fundamental Physics, Chalmers University of Technology, Gothenburg, Sweden}
\affiliation{CERN, EP-Division, CH-1211, Geneva 23, Switzerland}
\author{K. Rykaczewski}
\affiliation{Physics Division, Oak Ridge National Laboratory, Oak Ridge, TN 37831, USA}

\begin{abstract}
The photon spectrum accompanying the orbital K-electron capture in the first forbidden unique
decay of $^{81}$Kr was measured. The total radiation intensity for the photon energies larger
than 50 keV was found to be $1.47(6) \times 10^{-4}$ per K-capture. Both the shape of the spectrum
and its intensity relative to the ordinary, non-radiative capture rate, are compared to
theoretical predictions. The best agreement is found for the recently developed model which
employs the length gauge for the electromagnetic field.
\end{abstract}

\pacs{23.20.Nx, 23.40.-s, 29.30.Kv}

\maketitle

%====================================================================
\section{Introduction}
%====================================================================

Radiative electron capture (REC) is a process in which one of orbital electrons is captured
by the atomic nucleus and in addition to an electron neutrino a photon is emitted \cite{bambynek}.
Because of the three-body character of the process, the energy spectrum of these photons is continuous,
reaching up to the maximum value of the decay energy $Q_{EC}$ reduced by the binding energy of the
captured electron in the daughter atom, $B(nl)$. Such a radiative process occurs with the probability
of the order of $10^{-4}$ with respect to the ordinary, radiation-less electron capture.

Early studies have shown, for numerous decays involving allowed nuclear transitions
(nuclear spin changes by $\Delta J=0,1$ with no parity change),
that the intensity and shape of the REC spectrum can be well understood
by assuming that radiation is emitted by a captured electron
(internal bremsstrahlung, IB) \cite{bambynek}. However, a measurement
of REC in case of the forbidden nuclear transition in $^{41}$Ca \cite{41Ca}
revealed a strong disagreement with existing models. Moreover, the most advanced
theoretical model of IB, developed by Zon and Rapoport for the arbitrary degree of
forbiddenness \cite{zrap,zon},
showed the largest deviation from the experiment. The EC decay of $^{41}$Ca belongs to
a category of first-forbidden unique ($1u$) transitions ($\Delta J=2$, $\pi_i \pi_f = -1$).
Unique transitions are of special interest, because in the probability ratio of radiative
to non-radiative decays, the nuclear matrix elements cancel out and the process should
be governed solely by electromagnetic interactions. For $^{41}$Ca, the total probability
of the REC process, per non-radiative decay, was found to be 6 times \textit{larger} than
the detailed prediction of the Zon and Rapoport model \cite{41Ca}.

To explain this finding, a hypothesis of the so called \textit{detour} transitions was
considered by Kalinowski \cite{detour,kalinowski} following an idea of Ref.\cite{ford}.
According to it, a large part of radiation is emitted by a nucleus in addition to
the IB mechanism. It was argued that such a nuclear contribution is
particularly significant in case of $1u$ decays, where the combination of an allowed $GT$
EC decay with a nuclear $E1$ gamma transition, may compete with the direct nuclear $1u$
transition accompanied by an IB photon. Apparently, for the case of $^{41}$Ca this hypothesis
could fully account for the missing intensity found by the experiment \cite{detour,kalinowski}.
Surprizingly, in another case of the $1u$ REC --- the decay of $^{204}$Tl \cite{kurcewicz} ---
an opposite situation was observed. The intensity of the measured REC spectrum was
found to be \textit{smaller} by a factor of 4 than the predictions of the Zon and Rapoport
model. Thus, no room for detour transitions was left, since their contribution is always
positive in the model of Kalinowski.

To solve this conundrum, Pachucki et al. have undertaken a new approach
to description of the REC process \cite{pachucki}. The main ingredient of the
proposed model was the description of the electromagnetic field in the \textit{length}
gauge \cite{lgauge} in contrast to the Coulomb gauge used in all previous
calculations. The key point is that although the predictions are gauge-invariant,
the particular \textit{length} gauge is strongly preferred for the actual calculation
in this case. One reason is a suppression of the nuclear contributions, which allows
to neglect the detour transitions. The new model was found to agree
very well with experimental spectra of $^{41}$Ca and $^{204}$Tl \cite{pachucki}.

In this paper, we report on the REC measurements for the third case
of $1u$ transitions --- the decay of $^{81}$Kr. This nucleus, with an intermediate
mass and atomic number, located between $^{41}$Ca and $^{204}$Tl, suits very well as
a testing case for the theory of radiative electron capture accompanying
forbidden decays. In addition, this is the last case known where the
radiation accompanying a first-forbidden unique ($1u$) decay can be measured for the pure
ground-state-to-ground-state transition, which makes it experimentally feasible.

\subsubsection{Decay properties of $^{81}$Kr}

The $7/2^{+}$ ground state of $^{81}$Kr decays with the probability of almost 100~\%
to the $3/2^{-}$ ground state of $^{81}$Br with the half-life of
$2.3 \times 10^5$~y \cite{DecayData}. This transition is of the first-forbidden unique type.
The decay energy $Q_{EC}$ amounts to ($280.8 \pm 0.5$)~keV \cite{DecayData}, so that the decay
can proceed only by an electron capture process. A small branch of the $^{81}$Kr decay
feeds the first excited state of $^{81}$Br at 276 keV with the spin/parity of $5/2^-$.
The branching for this first forbidden non-unique transition was measured by Axelsson et al. to
be $3.0(2) \times 10^{-3}$ \cite{Axelsson}. However, the decay energy for this weak branch
equals only 4.7 keV and is much smaller than the K-electron binding energy in a bromine atom
which amounts to $B_{K}=13.47$~keV. Therefore, the transitions to the excited state
in $^{81}$Br proceed only by electron captures from higher shells. In turn, the K-electron
capture can occur only in a ground-state to ground-state transition. Thus, by a condition
of coincidences with Br KX-rays one can select pure $1u$ K-capture transitions between
ground states of $^{81}$Kr and $^{81}$Br. The maximum energy available for REC photons in the
K-capture is given by $q_K=Q_{EC}-B_K$, which leads to $q_K=267.3$~keV.

%====================================================================
\section{Experimental technique}
%====================================================================

\subsection{$^{81}$Kr source}

The activity of $^{81}$Kr was collected at CERN-ISOLDE \cite{Kugler}, using a proton beam of 1 GeV energy that induced spallation reactions in a $^{93}$Nb production target. The reaction products were ionized in a surface ion-source, accelerated to a kinetic energy of 50 keV and subsequently mass separated by means of the GPS separator tuned for the transmission of the particles with $A = 81$. The selected products were implanted in an aluminum catcher foil of 10 mg/cm$^2$ thickness. Krypton ions, although produced in the reaction, are not ionized and extracted from the source with any significant efficiency. However, rubidium is readily ionized with high yield and it was thus primarily $^{81}$Rb ions that were implanted, with some additional component of $^{81}$Sr. These implanted ions, having the half-life of 4.58 h and 22.2 m respectively, eventually decayed to $^{81}$Kr forming the source of interest.

To avoid sputtering of already implanted atoms by the impinging beam, the position of the foil was changed. After the irradiation, the foil was left in the chamber for a few days to let the short-lived precursor activity decay out. The final source material was collected on two foils in two irradiation sessions, in 1998 and in 2000, yielding six active spots, each having the area of a few mm squared.

Two longer-lived contaminants were found in the source, stemming from incomplete suppression of neighbouring isobars in the mass separator: $^{82}$Sr ($T_{1/2} = 22.5$~days) and $^{83}$Rb ($T_{1/2} = 86.2$~days). Their gamma ray activity, in particular Kr- KX rays, makes the $^{81}$Kr REC measurements very difficult. Instead of undertaking any chemical purification of the source, we decided to wait few years until these contaminants have decayed. The activity of $^{81}$Kr and of the contaminants was monitored. The gamma and KX spectra of collected samples, measured in 2001 and 2003, are shown in Figs. 1 and 2.

\begin{figure}[ht]
\begin{center}
    \includegraphics[width=1.0\linewidth]{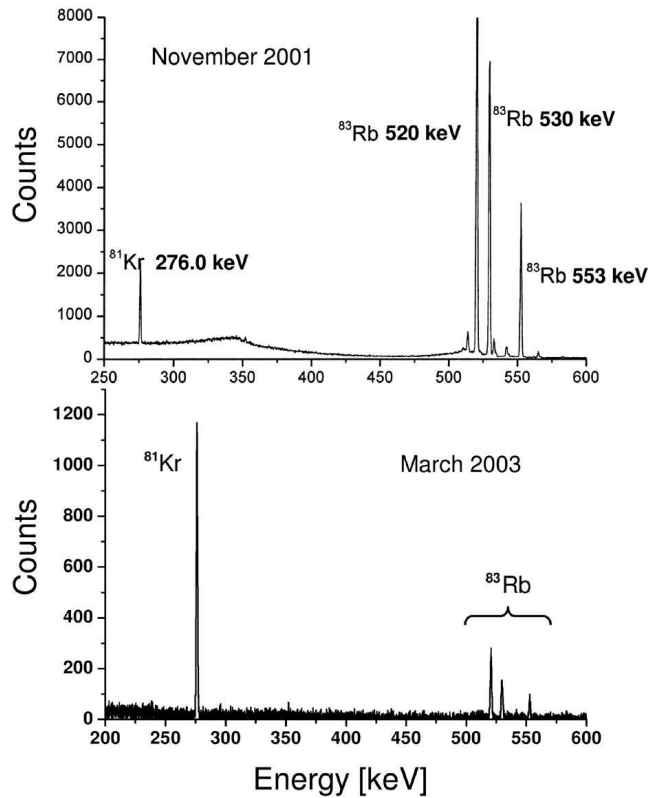}
    \caption{Singles gamma-ray spectra of the $^{81}$Kr source taken in November 2001 (upper
    panel) and in March 2003 (lower panel). Gamma transitions following decays of contaminant
    $^{83}$Rb atoms are seen as well as a 276 keV line from a decay of $^{81}$Kr to the 5/2$^{-}$
    excited state in $^{81}$Br.}
\end{center}
\end{figure}

\begin{figure}[ht]
\begin{center}
    \includegraphics[width=1.0\linewidth]{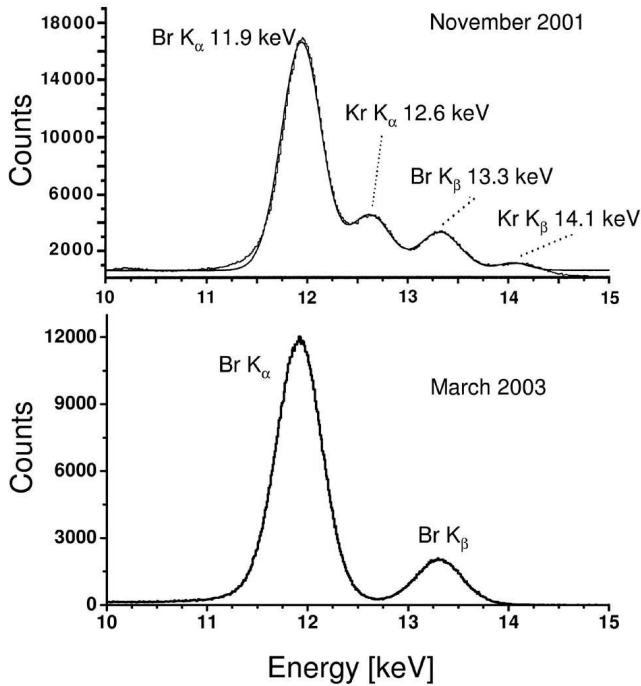}
    \caption{Singles KX-ray spectra of the $^{81}$Kr source taken in November 2001 (upper
    panel) and in March 2003 (lower panel). In the upper spectrum contaminant Kr KX-rays
    originating from decays of $^{83}$Rb can be seen.}
\end{center}
\end{figure}

Finally, in 2005 the influence of $^{83}$Rb was found to be negligible.
The active spots were carefully cut out from the catcher foils and placed on a
perspex disk 1 mm thick and 2 cm in diameter and covered tightly with a thin
mylar foil. The number of $^{81}$Kr atoms in the final
source was measured to be $4.1(2) \times 10^{15}$.

\subsection{Photon spectrometry}

Measurements were performed in the Institute of Experimental Physics
at the Physics Department of University of Warsaw. To select the K-capture component,
the REC spectrum was measured in coincidence with the Br-KX rays. REC photons
were recorded with an ORTEC GMX 45\% Ge detector. The X-rays were measured by means
of an ORTEC LOAX spectrometer. Both detectors had 0.5 mm thick beryllium windows.

The source was mounted between the two detectors in a close face-to-face
geometry, with the surface covered with mylar foil directed towards the X-ray detector.
The whole setup was placed inside a lead shielding of 10 cm thickness.
The inside of the shielding was covered with 1.2 mm thick cadmium sheets
and 8 mm thick copper plates.

A single DGF-4C CAMAC module \cite{XIA} was used to process electronic signals from
both detectors. The preamplifier outputs were directly connected to the two inputs
of the DGF-4C module. Each input signal was digitized with a 40 MHz frequency followed
by a real-time digital signal processing which included time stamping. Output data were
read out by a PC computer and stored on a hard disk. In the off-line analysis, the
coincidence relationships between signals could be restored with help of the time stamps.
In addition, a 1 Hz pulser was connected to the test inputs of both preamplifiers and to
a free running scaler. Comparison of the scaler readings with the number of counts
recorded by the DGF-4C module allowed to estimate and to monitor the dead time of the
acquisition system.

The long coincidence sessions (1-2 days) were alternated with short (ca. 10 min.) singles
measurements necessary for the KX-rays intensity monitoring.

Two series of measurements were carried out. In the first one, in 2005, the total time
of $\gamma$-KX coincidence runs amounted to 925.7 h ($\simeq 38.6$~days).
The average rate of singles Br-KX rays was determined to be $107.7(5)$ counts/s.
After all runs, the $^{81}$Kr source was removed and the background was measured in
the same coincidence conditions for 113 h.

In the second series of measurements, in 2007/2008, the $\gamma$-KX coincidences were
collected for total of 78 days. The average singles intensity of Br-KX rays over this
period was found to be $95.1(5)$ counts/s. This time, however, for reasons explained
in the following, the $^{81}$Kr runs were alternated with background $\gamma$-KX
coincidences. The total time of background coincidence measurements was 71 days.
The total time of the second experiment amounted to 190 days.

\subsection{Calibrations and corrections}

The calibrations of both Ge detectors, including the efficiency calibration of the GMX
detector, were performed with help of standard sources: $^{57}$Co, $^{60}$Co,
$^{137}$Cs, $^{152}$Eu, $^{203}$Hg, and $^{241}$Am. In order to avoid summing effects,
the full-energy peak efficiency of the GMX detector was measured in a far geometry
with all sources and
in the close geometry, corresponding to the $^{81}$Kr source position,
only with $^{241}$Am, $^{203}$Hg, and $^{137}$Cs, in which no summing occurs.
Then, the far-geometry curve was scaled to the close-geometry position based on the
efficiency ratios determined with the latter sources. The finally determined efficiency
of the GMX detector at 100 keV amounted to about 14\% for both runs of $^{81}$Kr
measurements.

The performance of the coincidence circuitry was tested with a $^{133}$Ba source measured
in a far geometry to avoid summing effects. From the decay scheme of $^{133}$Ba the
probabilities of $\gamma$-line emission per K-capture were determined for the lines
at 81 keV, 276 keV, 302 keV, 356 keV, and 384 keV. These values were compared
with the measured $\gamma$-KX coincidences yielding the coincidence efficiency
consistent with 100\% within error bars of a few percent for all tested energies.

Since the REC spectrum of photons is continuous, it has to be corrected for the
Compton scattering. To estimate the magnitude and shape of the corresponding
correction, the response of the GMX detector to mono-energetic radiation had to be
determined. First, the response was measured with help of sources: $^{241}$Am,
$^{57}$Co, $^{203}$Hg, and $^{137}$Cs. Then, the spectra were compared with
Monte-Carlo simulations performed with the GEANT package \cite{Geant}. The effective
dimensions of the detection set-up were adjusted to obtain a satisfactory
agreement between the measured and simulated spectra. This allowed to calculate
the detector response to the radiation of arbitrary energy within the range of interest.

\subsection{Background}

Selection of the K-component of the REC spectrum requires coincidences with KX-rays
of the daughter atom. Usually, such condition practically removes any background
contribution to the measured gamma spectrum. However, in case of $^{81}$Kr, the
energy values of Br-KX rays, 11.9 keV and 13.3 keV, unfortunately happen
to overlap with LX-ray energies of heavy elements, like protactinium and thorium.
Since some radioactive isotopes of these elements belong to natural
radioactivity chains, traces of them appear in the surroundings, mainly in lead bricks.
Coincidence conditions result in a selection of those gamma rays from their decays
which are coincident with transitions strongly converted on the L-shell.
An example of background gamma spectrum, measured without the $^{81}$Kr source
but with the same coincidence conditions, in particular
with a gate on Br-KX energy range in the LOAX detector, is shown in Fig. 3.

\begin{figure}[ht]
\begin{center}
    \includegraphics[width=1.0\linewidth]{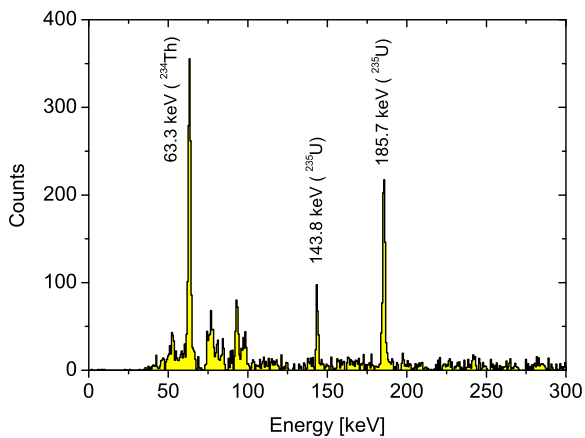}
    \caption{The background gamma spectrum gated by radiation in the Br-KX ray energy
     range detected in the LOAX detector. The total measurement time was 70 days. }
\end{center}
\end{figure}

The strongest $\gamma$ lines present in Fig. 3 are identified as emitted by members of uranium-radium and
actinium natural radioactivity chains, starting from $^{238}$U and $^{235}$U, respectively.
For example, a strong line at 63.3 keV is emitted from $^{234}$Pa after the
$\beta ^-$ decay of $^{234}$Th (itself a decay product of $^{238}$U).
This line is seen because of coincidence with an E2 transition of 29.5 keV
which has a large L-shell conversion coefficient in protactinium.
Another L-converted transition of 20.0 keV (M1+E2) in $^{243}$Pa,
is in coincidence with the 92.4 keV gamma line which can also be seen in the spectrum.
Two other strong lines, at 143.8 keV and 185.7 keV, correspond to transitions
between states in $^{231}$Th fed by $\alpha$ decay of $^{235}$U.
Both are coincident with a 19.6 keV M1+E2 transition
which is strongly L-converted.

The contribution of natural background to the REC spectrum was identified
after the first series of measurements when the coincidence run without
the $^{81}$Kr source was taken only for 113 h. Therefore, in the second series,
the coincident background runs were alternated with REC runs, so that the background
spectrum with much higher statistics was accumulated.

%====================================================================
\section{Results}
%====================================================================

\subsection{REC spectrum}

The K-component of the REC spectrum of $^{81}$Kr was determined separately from both
measurement series by applying the following procedure. First, the spectrum of time
differences between $\gamma$-ray and KX-ray events was created. The peak in this
spectrum allows to discriminate the true KX-$\gamma$ coincident events from a flat background
representing random coincidences. Then, by appropriate gating on the X-ray coordinate
to select Br-KX rays and on the spectrum of time differences, the spectrum of $\gamma$ rays
coincident with Br-KX rays was constructed for all runs measured with the $^{81}$Kr source.
Exactly the same gating procedure was used
to determine the $\gamma$ spectrum of background coincidences, discussed in the previous
section. After normalization, the background spectrum was subtracted from
the former one yielding the raw REC spectrum of $^{81}$Kr accompanying the K-shell
electron capture. As an example, the spectra from the second experiment are shown in Fig. 4.
The total number of counts in the REC spectrum is 16800, which
corresponds to about 9 coincident events per hour.

\begin{figure}[ht]
\begin{center}
    \includegraphics[width=1.1\linewidth]{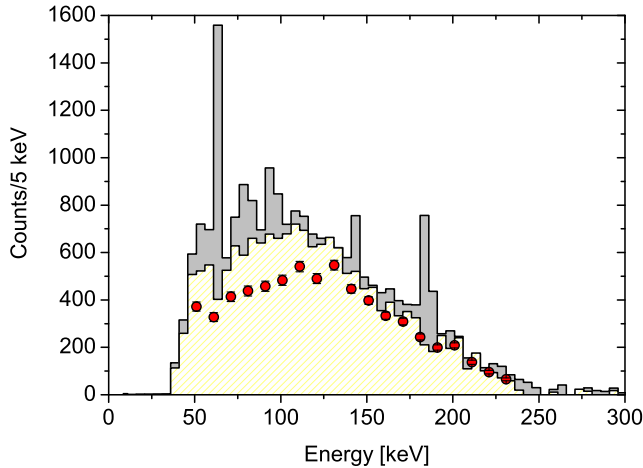}
    \caption{(Color online) Gamma energy spectrum coincident with Br-KX rays 
             from the second series of
             measurements. The upper histogram was
             extracted from runs with $^{81}$Kr source. The contribution of background
             lines can be seen. The lower histogram shows the result of background
             subtraction. The circles filled in red illustrate the result of Compton correction
             and represent the raw K-component of the $^{81}$Kr REC spectrum.  }
\end{center}
\end{figure}

In the next step, the correction for Compton scattering is introduced. Calculated contributions
of each $\gamma$-energy bin are subtracted one by one, starting from the high-energy end
(a peeling-off method). This correction is largest at low energy where contributions
from all higher-energy bins add up, and amount to about 30\%. The corrected REC spectrum is
shown in Figure 4 by points with the error bars.

Finally, the absolute normalization of the REC spectrum is made. Each bin of the Compton-corrected
REC spectrum is divided by the corresponding full-energy-peak efficiency of the GMX detector and
by the total number of Br-KX rays recorded by the LOAX detector during the all $^{81}$Kr
coincidence runs. This number was determined from the singles KX-ray measurements and includes
corrections for the dead-time in both the singles and coincidence runs. Such a procedure
yields the probability distribution of the REC photon energy normalized to ordinary, non-radiative K-capture rate. This final REC energy spectrum can be directly compared to theoretical predictions.
The resulting distributions from two series of measurements are presented in Figure 5.

\begin{figure}[ht]
\begin{center}
   \includegraphics[width=1.0\linewidth]{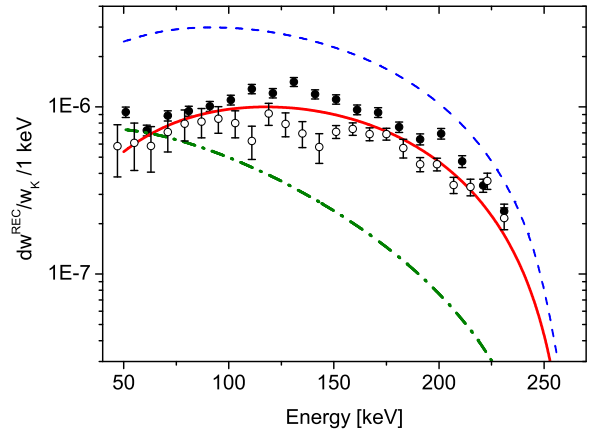}
    \caption{(Color online) The intensity of the K-component of the REC spectrum of $^{81}$Kr per ordinary, non-radiative K-capture and per 1 keV. Open and closed circles represent the first and the second series of measurement, respectively. The solid line shows the prediction of the model by Pachucki et al. \cite{pachucki}, while
    its Coulomb-free limit is represented by the dashed line. The dash-dotted line illustrates the prediction
    of the Zon and Rapoport model \cite{zrap,zon}.}
\end{center}
\end{figure}

\subsection{Theoretical models}

In a very general way, the probability of radiative electron capture from the 1S state (K-capture)
in which a photon in the energy range $(k, k+dk)$ is emitted, per ordinary, non-radiative K-capture
can be written as \cite{bambynek}:
\begin{equation}
\frac{dw^{REC}_{K}(k)}{w_{K}} = \frac{\alpha}{\pi(m_e c^2)^2}
\frac{k(q_{K}-k)^2}{q^2_{K}} R_{K}(k) \, dk,
\label{eq:ms}
\end{equation}
where $\alpha$ is the fine structure constant, $m_e c^2$ is the electron rest energy, $q_K$
is the photon end-point
energy, and the dimensionless function $R_{K}(k)$ is the shape factor of the spectrum.
In the simplest Coulomb-free,
non-relativistic approximation, valid for allowed nuclear transitions $R_K (k) \equiv 1$ \cite{MS}.

In a more appropriate theoretical approach which takes into account both relativistic effects and
the influence of the Coulomb field, and which is valid for first-forbidden unique
nuclear transition, the shape factor is given by:
\begin{equation}
R^{1u}_{K}(k) = \left(1-\frac{k}{q_{K}}\right)^2
      R^{\,(1)}_{K}(k) +
      \left(\frac{k}{q_{K}}\right)^2
      R^{\,(2)}_{K}(k),
\label{eq:ra}
\end{equation}
where the functions $R^{\,(1)}_{K}$ and $R^{\,(2)}_{K}$ describe the effects of the Coulomb interaction
between the nucleus and the radiating electron. Usually, these functions have to be calculated
numerically for each specific case.

The first advanced description of REC in forbidden transitions was provided by Zon and
Rapoport \cite{zrap,zon}, who extended a framework for allowed decays, developed previously
by Glauber and Martin \cite{GM,MG}, and generalized it to nuclear transitions of any order of forbiddenness.
In this model both functions, $R^{\,(1)}_{K}$ and $R^{\,(2)}_{K}$, approach
unity in the Coulomb-free limit ($Z \rightarrow 0$). The model of Zon and Rapoport was found to fully
reproduce the earlier results of Glauber and Martin for allowed decays which were quite well confirmed
experimentally \cite{bambynek}.
However, both the full and the Coulomb-free versions of this model failed to describe
the REC spectrum measured in case of $1u$ decays in $^{41}$Ca and $^{204}$Tl \cite{41Ca,kurcewicz}.

This observation motivated Pachucki et al. to reconsider the problem of REC in
forbidden decays following a different approach \cite{pachucki}. The important conclusion from
that work was that, although final results must not depend on a particular choice of
gauge for the electromagnetic field, the Coulomb gauge applied by previous authors has rather
unfortunate consequences when, as exemplified in the work of Zon and Rapoport, the approximations
may later result in diverging terms. In addition, in the Coulomb gauge one has to include contributions from
nuclear degrees of freedom. In turned out that the adoption of a different gauge for the electromagnetic field, so called \emph{length} gauge, simplifies calculations considerably and avoids, in fact, the difficulties mentioned before. In particular, it was demonstrated that in the \emph{length} gauge
the contribution form the nuclear (detour) transitions can be neglected \cite{pachucki}.

The results of Pachucki et al.
confirmed all previous predictions for allowed transitions as well as the detailed form of the function
$R^{\,(1)}_{K}$, appearing in eq. (2). However, a different form
was derived for the second function --- $R^{\,(2)}_{K}$, which is the one affected by a diverging term in
the model of Zon and Rapoport. In particular, the Coulomb-free limit for this function was found
to be \cite{pachucki}:
\begin{equation}
\lim_{Z\rightarrow 0} {R^{\,(2)}_{K}(k)} =  1+\frac{m_e c^2}{k} + 2 \left( \frac{m_e c^2}{k} \right)^2 ,
\end{equation}
which obviously differs from 1.

The model of Pachucki et al. was found to be in almost perfect agreement with measured spectra
in $^{41}$Ca and in $^{204}$Tl \cite{pachucki}. Its prediction for the present case of $^{81}$Kr
is plotted in Figure 5 with the solid line. The result of the Coulomb-free limit and of the
Zon and Rapoport model are also shown.

\subsection{1S shape factor}

The shape of the REC spectrum, contributions of different terms, and comparison with
data can be shown in detail by plotting the dimensionless shape factor of the spectrum, $R_{K}(k)$.
The experimental shape factors were extracted from the determined REC spectra with help of
Eq. (1). The values obtained are shown in Figure 6 which contains also theoretical predictions.

As seen in Figs. 5 and 6, there is a systematical difference between results obtained from two
experiments. The collected statistics in the first series was smaller which is partly reflected
in larger errors. In addition, the correction for background was less accurate in the first
experiment, which could result in a systematical shift. Since the investigated effect is small
and the rate of coincidences was very low, a proper control of possible systematical
effects was difficult. We note however, that each spectrum taken
separately would lead to the same conclusion. The prediction of Pachucki et al. reproduces quite
well both experimental spectra, in contrast to the model by Zon and Rapoport.

\begin{figure}[ht]
\begin{center}
   \includegraphics[width=1.0\linewidth]{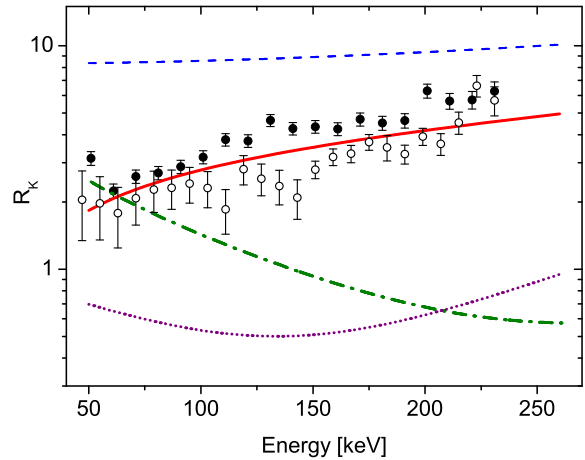}
    \caption{(Color online) The shape factor for the K-component of the 
    first forbidden unique REC in $^{81}$Kr. Open and closed circles represent the first and the
    second series of measurement, respectively.
    The meaning of theoretical lines is the same as in Fig. 5, while 
    the additional dotted line shows
    the Coulomb-free version of the Zon and Rapoport model.}
\end{center}
\end{figure}

\subsection{Total intensity}

The total probability of the K-component of the REC spectrum per non-radiative K-capture,
calculated by integrating the spectrum from 50 keV to the end-point, is given in Table~1 together
with values obtained from theoretical models.
The weighted average of the total intensity from two measurements equals to $1.47(6) \cdot 10^{-4}$ per K-capture.

\begin{table}
\caption{The total probability for the K-component of REC per K-capture in $^{81}$Kr, integrated
from 50 keV to the spectrum end-point, in units of $10^{-4}$ obtained from the first (a) and
the second (b) experiment. The results for a full calculation, as well as the Coulomb-free version (CF), are given for both theoretical models.}

\vspace{0.5\baselineskip}
\begin{centering}
\begin{tabular}{ccccc}
\hline
\multicolumn{1}{c}{Experiment} &
\multicolumn{2}{c}{Pachucki et al.} &
\multicolumn{2}{c}{Zon and Rapoport} \\
\multicolumn{1}{c}{ } &
\multicolumn{1}{c}{Full} &
\multicolumn{1}{c}{CF} &
\multicolumn{1}{c}{Full} &
\multicolumn{1}{c}{CF}\\
\hline
   a) 1.2(1) &  1.37 & 4.00 & 0.60 & 0.26 \\
   b) 1.64(8) &       &      &      &      \\
\hline
\end{tabular}
\end{centering}
\end{table}

The value of total intensity from both measurements, as well as their average, are closest to the
prediction of Pachucki et al. It is interesting to note the large difference between the Coulomb-free
limits of both theories. In case of Pachucki et al., this approximation overestimates the spectrum
intensity. The inclusion of Coulomb and relativistic effects reduces the intensity to the value
which almost perfectly agrees with the experiment. The model of Zon and Rapoport behaves in the opposite
way. The Coulomb-free limit lies below the full calculation, which in turn is still much lower than the experimental values.

%====================================================================
\section{Conclusions}
%====================================================================
We have measured the spectrum of photons accompanying the first-forbidden unique electron capture
decay of $^{81}$Kr. By detecting $\gamma$-rays in coincidence with Br-KX rays we have
selected the pure K-component of radiative electron capture (REC) taking place
between ground states of $^{81}$Kr and $^{81}$Br. From the measured
spectra we have extracted the absolute probability of REC emission per normal, non-radiative
K-electron capture as a function of photon energy, as well as the shape factor of the photon
spectrum. Both the shape and intensity of the REC spectrum are found to be well
reproduced by the recent theoretical model of Pachucki et al. \cite{pachucki}, while the
strong disagreement is found with predictions of the old model by Zon and Rapoport \cite{zrap,zon}.
Thus, $^{81}$Kr is the third case displaying REC in a \emph{1u} nuclear transition,
in addition to $^{41}$Ca and $^{204}$Tl, which strongly supports the approach taken by
Pachucki et al. In contrast, the model of Zon and Rapoport fails to describe correctly
the probability and shape of the REC spectrum in all these three cases.

The important feature of the first-forbidden unique decays is that they
are governed predominantly by a single nuclear matrix element.
In the probability ratio of the radiative capture to the ordinary,
non-radiative one, which is determined in experiment, this nuclear matrix element
cancels out. Thus, the result should not depend on detailed knowledge of nuclear wave
functions and should be fully determined by the electro-weak sector alone.
An important consequence is that the theoretical models discussed do not
contain any adjustable parameters. The predictions are fully determined by
the mass and atomic numbers of the decaying nucleus and
by the maximal photon energy, related to the decay energy $Q_{EC}$.

The main difference between the two theoretical models discussed is the
selected gauge of electromagnetic field. While the
Zon and Rapoport use the \emph{Coulomb} gauge, Pachucki et al. apply
the \emph{length} gauge.
Our conclusion is that the latter is preferred. The reason is that
calculations in the \emph{length} gauge are technically
simpler and do not require additional approximations.
Also, Zon and Rapoport introduced approximations which were not
valid and led to a diverging term which was found to be responsible
for the final failure in comparison with the experiment.
Moreover, in the \emph{length} gauge the contribution from nuclear
degrees of freedom to the emitted radiation can be shown to be negligible,
while this is not true in case of the Coulomb gauge.

The $1u$ decays of $^{41}$Ca, $^{81}$Kr, and $^{204}$Tl are the only known
where the pure transition between ground states can be selected. Thus, only
in such cases, the weak radiative branch can be determined with sufficient
accuracy. Additional tests of this alternative description of REC are
in principle possible with transitions of second-forbidden non-unique
type (\emph{2nu}: $\Delta J=2$, $\pi_i \pi_f = +1$). To this class
belong $^{59}$Ni and $^{137}$La in which K-component of REC spectrum
was measured \cite{59Ni,137La}. The results obtained for $^{59}$Ni
disagree with Zon and Rapoport predictions, while the spectrum for
$^{137}$La could not be shown to contradict them. However, in case
of \emph{2nu} decays, the REC spectrum depends on one additional parameter
which represents a ratio of nuclear matrix elements. Since these
matrix elements are not known, and are difficult to estimate,
this parameter adds an extra degree of freedom. In
consequence, the comparison between an experiment and the the theory
is not as unambiguous as in case of \emph{1u} transitions.
Nevertheless, in the future we plan to compare the measured spectra
of $^{59}$Ni and $^{137}$La with predictions of the new model
of Pachucki et al.

\section*{Acknowledgements}

%acknowledgements
We are grateful to the CERN-ISOLDE staff for assistance and patience during
irradiations to produce the samples of $^{81}$Kr.

\end{document}